\documentclass[12pt]{article}
\usepackage{graphicx}
\usepackage{authblk}
\usepackage[hidelinks]{hyperref}
\usepackage{orcidlink}
\usepackage[authoryear,round]{natbib}

\title{\vspace{-4.5em}Easy Reads: A Python program for making Scientific Papers on arXiv more Reader Friendly and Accessible}

\author[1,2,3]{\hspace{+0.8em}Vishal Verma~\orcidlink{0009-0003-9557-986X}}

\affil[1]{Department of Astrophysics, American Museum of Natural History, Central Park West and 79th Street, NY 10024-5192, USA}
\affil[2]{The Graduate Center of the City University of New York, 365 Fifth Avenue, New York, NY 10016, USA}
\affil[3]{Department of Physics and Astronomy, Lehman College of the CUNY, Bronx, NY 10468, USA}
\date{\vspace{-1.2em}\small\today}

\begin{document}

\maketitle

\section{Summary}

Scientific papers are frequently dense and characterized by features such as small fonts and line spacing, double columns of text, and tightly arranged figures. While these features make papers more compact, they can hinder readability, make them less accessible, and can strain the reader. arXiv is a premier open-access repository for scientific papers across different fields and is used extensively by researchers, including those in the physics and astrophysics communities.

Easy Reads is an automated, end-to-end, open-source Python program that helps address the stated challenge by making papers from arXiv more reader-friendly and accessible. Easy Reads can automatically fetch a paper from arXiv via its URL and work with the source TeX file to allow custom formatting of the paper features, primarily the font size, and the number of columns used. The main goal of Easy Reads is to facilitate ease of reading of scientific papers.
 
\section{Statement of need}

The small fonts, double columns of text, and tightly arranged figures in scientific papers are examples of features that were historically chosen by publishers to account for economic factors and optimize page count when printing papers. While several journals have now gone completely online, research papers still inherit formatting standards from before. This negative impactd the reader. While zooming in on the document helps increase the font size, the process is not seamless when navigating a document, often requiring the reader to zoom in and out. Zooming in also does not intrinsically change the document; thus, a printed version will contain the same number of pages and show no improvement.  Easy Reads aims to tackle factors that can make scientific papers difficult to read, which we discuss in this section.

\subsection{Digital Reading and the Effect of Font Size}

In an increasingly digital world, researchers spend substantial amounts of time reading scientific papers on screens. Prolonged screen use has been associated with a range of symptoms collectively referred to as Digital Eye Strain (DES), including dry eye, visual discomfort, and fatigue. The strain over time can also lead to musculoskeletal conditions such as pain in the neck, shoulder or head and potentially impact productivity; and a small font size is an additional overhead \citep{gowrisankaran_computer_2015, rosenfield_computer_2011}. Small font sizes can increase fixation period (duration where the eye remains fixed on a word or region of text during reading), cause a reduction in blinking rate and these effects are pronounced for cognitively demanding works like scientific papers \citep{rosenfield_computer_2011}.  Readers with preexisting visual, ocular, musculoskeletal, or headache-related conditions may have their symptoms exacerbated. People who are older and those with vision issues are particularly sensitive to the font size and formatting \citep{rubin_effect_2006}. Font size below a threshold also negatively impacts reading speed \citep{atilgan_reconciling_2020}.  Some studies suggest that larger font size words may be remembered better \citep{halamish_effect_2018,chang_association_2022}. Thus, overall, font size is a key factor to be considered as part of the reading experience.

\subsection{Number of Columns}

The two column layouts in papers could be disruptive for some readers. They can introduce additional navigation demands and alter fixation patterns \citep{shrestha_eye_2008}. Single-column layouts have been have been reported to improve reading speed in some scientific-journals and screen-reading contexts, and have been recommended for accessible document design. Easy Reads provides an option to switch the document to a single-column layout.

\subsection{Print versus Screen Reading}

Because Easy Reads works with the source TeX file to modify the font size of a paper, the resulting printed version will contain more pages than the original. There are some advantages of a printed paper versus its digital counterpart. Despite the prevalence of digitized versions of papers, when academic materials are lengthy, students still tend to prefer reading on printed paper \citep{rose_phenomenology_2011}. Academic documents also require a higher level of concentration and cognitive effort. Some studies, for instance, \citet{durant_future_2015} and \citet{jeong_advantages_2021} suggest that, for complex and lengthy papers, the printed medium may aid a deeper sense of understanding. Additionally, a printed document has the natural advantage of being free of the distractions that may sometimes accompany a digital medium.

\subsection{Selecting an Optimum Font Size}

The font size adopted across scientific journals varies and may also depend on whether a paper is in the preprint or published stage. As an example, the font size for some astrophysics journals can be about $10$ points or lower. Several studies have explored an ideal range of font sizes suitable for reading on screens and in print. \citet{banerjee_selection_2011} found that a font size of $14$ points for computer screens was suitable for readers, and that pupil diameter, fixation duration, and gaze duration were minimized at this font size. \citet{rello_make_2016} found that an even larger font size is especially suitable for text-heavy websites. \citet{rubin_effect_2006} observed that, for people with vision issues, changing the font size in print from $10$ points to $14$ or higher would be helpful and significantly increase their participation. Easy Reads defaults to a font size of $12$ points, but this can be changed as desired.

Given the impact that formatting can have on readability, accessibility, visual-strain-related effects, and the overall reading experience, the author's view is that it may be worthwhile for scientific journals to consider offering customizable options for paper outputs, especially in PDF format.

\section{Software Design and Usage}

Easy Reads offers users two key customization options: adjustable font size of the text of the main body and an optional single-column formatting for the entire paper. The line spacing and margins are computed automatically for optimal readability but can be manually tuned if desired. 

\subsection{Input Requirements}

Easy Reads is designed to be user-friendly and requires three main inputs to the primary script (\texttt{main\_easy\_reads.py}):

\begin{enumerate}
    \item \textbf{arXiv URL}: The paper's abstract page, in the format: \begin{center}
        \texttt{https://arxiv.org/abs/XXXX.YYYYY}
    \end{center}, where \texttt{XXXX} and \texttt{YYYYY} are unique numerical identifiers)
    \item \textbf{Font size}: Base font size for the text of the body (default: 12pt)
    \item \textbf{Layout mode}: Option to enable single-column formatting (default: preserve original column layout)
\end{enumerate}

\subsection{Processing Pipeline}

Easy Reads automates the following workflow:

\begin{enumerate}
    \item \textbf{Source extraction}: Given the arXiv URL, retrieves the LaTeX source files located at \begin{center}
\texttt{https://arxiv.org/src/XXXX.YYYYY}
    \end{center}
    \item \textbf{Download and extraction}: Downloads the source file (typically in the \texttt{.tar.gz} format) to a \texttt{Downloads/} folder (located in the same directory as the  \texttt{main\_easy\_reads.py} file) and extracts its contents.
    \item \textbf{Modification}: Locates and edits the main \texttt{.tex} file with the specified settings.
    \item \textbf{Compilation}: Compiles the modified LaTeX to PDF and saves the output to a \texttt{Formatted Papers/} folder.
    \item \textbf{Naming}: Appends a customizable suffix to the original filename (default: \texttt{\_formatted}, e.g., \texttt{9001.10000v1\_formatted.pdf}) to distinguish it from the original.
\end{enumerate}

\subsection{Two Usage Modes}

Easy Reads provides two approaches for customizing papers. The recommended method is to use the command-line interface (CLI). The other option is to edit the parameters directly in the \texttt{main\_easy\_reads.py} file.

\subsubsection{Mode 1: Command Line Interface (Recommended)}

Pass parameters directly via CLI arguments via a single line for flexible control.
\vspace{0.2cm}

\textbf{Basic usage:}
\begin{verbatim}
python main_easy_reads.py --url https://arxiv.org/abs/XXXX.YYYYY
\end{verbatim}
This would download the source file to the \texttt{Downloads/} , unzip its contents, apply a font size of $12$ (default, since it is unspecified), and produce a PDF output in the \texttt{Formatted Papers/} folder. The number of columns will be the same as the original file, since this argument is unspecified. By default an output suffix  \texttt{\_formatted}  is appended to the name of the paper but this can be modified.
\vspace{0.2cm}

\textbf{Allowed CLI arguments} (listed by order of importance):

\begin{center}
{\footnotesize
\begin{tabular}{|l|p{4.5cm}|p{3cm}|}
\hline
\textbf{Argument} & \textbf{Description} & \textbf{Default} \\
\hline
\texttt{--url} & arXiv paper URL (required) & (none -- required) \\
\hline
\texttt{--font-size} & Base font size in points & \texttt{12} \\
\hline
\texttt{--single-column} & Enable single-column format & \texttt{False} \\
\hline
\texttt{--single-column-margin} & Custom margin width in inches for single-column mode & \texttt{None}  (auto-scales with font size)\\
\hline
\texttt{--baseline} & Line spacing in points & 1.2 $\times$ font size \\
\hline
\texttt{--output-suffix} & Suffix for output filename & \texttt{\_formatted}\\
\hline
\end{tabular}
}
\end{center}
 \vspace{0.5cm}

The single-column margin is set to 1.5\,in.\ at 12\,pt and auto-scales with the font size, but this can be tuned.

 \vspace{4cm}

\textbf{Example commands:}
\begin{verbatim}
# Larger font for easier reading 
# (Note:Default of 12 points generally increases the fontsize) 
python main_easy_reads.py --url https://arxiv.org/abs/XXXX.YYYYY \
    --font-size 14

# Single-column layout with auto-scaled margins
python main_easy_reads.py --url https://arxiv.org/abs/XXXX.YYYYY \
    --single-column
    
# Adding a Custom output suffix
python main_easy_reads.py --url https://arxiv.org/abs/XXXX.YYYYY \
    --output-suffix _easy
\end{verbatim}

The first example above sets the font size of the paper to 14 . The second one activates the single column mode. The last appends a \texttt{\_easy} to the output (default would be \texttt{\_formatted}) , for instance if the original name is \texttt{9001.10000v1} the generated output will have the name \texttt{9001.10000v1\_easy}, clearly distinguishing it from the original.

\subsubsection{Mode 2: Direct Code Editing}

For users who prefer to work with the main script \texttt{main\_easy\_reads.py} directly:

\begin{enumerate}
    \item Open \texttt{main\_easy\_reads.py} in your preferred editor.

    \item Locate the settings section at the bottom of the script and modify the following variables as needed:\\
    \texttt{URL}, \texttt{FONT\_SIZE}, \texttt{BASELINE},
    \texttt{SINGLE\_COLUMN}, \texttt{SINGLE\_COLUMN\_MARGIN},
    and \texttt{OUTPUT\_SUFFIX}.
    
    \item Run the script:
\begin{verbatim}
python main_easy_reads.py
\end{verbatim}
\end{enumerate}
\section{Related Projects}

The currently existing alternatives are in practice limited in scope or designed for different use cases. Recently, arXiv introduced an experimental HTML version for research papers, which represents a promising step toward more accessible scientific reading. However, this feature is currently available only for a subset of papers, primarily newer submissions and papers from the recent few years, with older content expected to be added over time. Furthermore, HTML based papers may not integrate seamlessly with reference-management workflows and annotation tools, such as Zotero, that many researchers rely on for highlighting, note-taking, and literature organization. Printed versions of HTML rendered papers may also exhibit formatting artifacts. Some publishers additionally provide papers in ePub format, a highly flexible option to allow readers to customize font size, layout, and other presentation settings. However the output quality varies across journals and ePub readers, and not every journal provides this option. More importantly, such options are generally limited to published papers and do not address the large body of preprints available through repositories such as arXiv. Thus, Easy Reads is uniquely positioned to tackle the aforementioned challenges.

\section{GitHub Repository and Future Work}

Easy Reads is maintained on GitHub at:
\begin{center}
\url{https://github.com/Curious-flow/Easy-Reads}
\end{center}

Easy Reads aims to be agnostic across academic disciplines. However, formatting compatibility and output quality may vary across journals due to journal-specific LaTeX practices and custom packages. Potential issues may include, but are not limited to, single-column formatting, margin and spacing inconsistencies, and the sizing of figures and equations.

Easy Reads is currently in its alpha release and remains a work in progress. If you encounter any difficulties, please report an issue on GitHub.

\vspace{0.2cm}

\subsection{Future Work}

Easy Reads is an independent project and is not affiliated with arXiv in any way. Future versions of Easy Reads may provide additional customization options, such as adjustments to the font sizes of titles, abstracts, and section headings, as well as the resizing of figures and tables.

\section{AI usage disclosure}

Large Language Models (LLMs) were used as coding assistants during the development of this project. The author reviewed and validated any generated code and assumes responsibility for the final work. 

\section{Acknowledgements}

The author acknowledges arXiv for its invaluable role in providing open access to scientific literature and, where available, source files that support learning and reproducible research. The author thanks the open-source community behind the LaTeX ecosystem, including MiKTeX and TeX Live, for providing the tools that make scientific writing and publishing accessible. The author extends their gratitude to Viviana Acquaviva for her encouragement during the development of this project. 

{\footnotesize
\begingroup
\renewcommand{\url}[1]{}
\providecommand{\urlprefix}{}
\renewcommand{\urlprefix}{}
\bibliographystyle{plainnat}
\bibliography{References}
\endgroup
}

\end{document}